# A polarization-Insensitive Broadband Achromatic Metalens with High Efficiency in Ultraviolet-C band

Hong Song Fang, Ly Ly Nguyen Thi and Shu-Chun Chu∗

*Department of Physics, National Cheng Kung University, No. 1, University Road, Tainan City 701, Taiwan*

*\*scchu@mail.ncku.edu.tw*

**Abstract:** A metalens is composed of an array of artificially designed meta-atoms which can manipulate the phase, polarization and amplitude of light making it an excellent component for wavefront modulation. In this study, we design an transmittive achromatic metalens (NA equals to 0.05) composed of sapphire substrate and cross-shape silica meta-atom array operating across the entire ultraviolet C (UVC) band (200–280 nm) for all incident polarization through numerical simulation. It is shown that the device achieves an average focusing efficiency of 75% with a focal shift of less than 10% across the bandwidth and successfully demonstrates that is polarization-free. The device is expected to perform effectively in ultraviolet imaging and hold great potential for space-based astronomical observations.

## 1. Introduction

Metasurfaces are planar optical elements formed by the periodic array of the engineered subwavelength meta-atoms on a supporting substrate. Each meta-atom, defined by its shape, geometric parameters and material properties, interacts with incident light to modulate its amplitude, phase, or polarization [1-5]. They offer several advantages, including compact size, light weight , and multifunctionality, enabling the realization of the highly miniaturized imaging systems [6-8]. Consequently, metasurfaces have been widely applied in various areas such as meta-lenses [9-11], meta-holography [12-14], meta-displays [15,16] and so on. By appropriately spatially arranging the meta-atoms to shape the desired wavefront, metasurfaces can operate as the ultrathin lenses, commonly referred to as metalenses.

Over the past few decades, conventional refractive optical elements have exhibited inherently strong wavelength dependence, which leads to severe chromatic aberrations. Chromatic correction is typically achieved by cascading multiple lenses made from different materials, which increases system size while still providing only a limited achromatic bandwidth [17]. Metalenses offer an alternative planar strategy: by selecting appropriate material platforms and carefully engineering the geometry of individual meta-atoms, both the phase and its dispersion can be simultaneously controlled, enabling broadband chromatic aberration correction within a single ultrathin element [18]. Moreover, metalens fabrication is compatible with mature semiconductor manufacturing technologies, such as complementary metal–oxide–semiconductor (CMOS) processes, allowing high-precision and large-scale production [19]. Therefore, dispersion-engineered metalenses not only overcome the severe chromatic aberrations inherent to conventional diffractive planar optical elements, but also demonstrate strong potential as compact and manufacturable alternatives to bulky multi-element lens systems.

So far, many broadband achromatic metalenses (BAML) have been successfully demonstrated, operating in visible [20,21], near-infrared [22], and mid-infrared [23,24] spectral regimes, with some designs even achieving performance across adjacent bands [26]. These devices are typically constructed from the meta-atoms with the different dispersion properties and dynamic phase, so that both the required phase and group delay (GD) can be satisfied simultaneously, sometimes with the aid of Pancharatnam–Berry (PB) phase [25] compensation.

More recently, several achromatic tunable or varifocal meta-devices have been proposed, including designs that extend the depth of focus, integrate polarization-dependent phase control, or place a metalens together with an additional polarization rotator to correct chromatic aberrations. However, research on achromatic metalenses operating in the ultraviolet (UV) and extreme ultraviolet (EUV) spectral regions remains rather limited, with reported efficiencies generally not reaching 75% [26-28]. This limitation primarily arises from stringent material constraints: at such short wavelengths, only a very small number of dielectric materials simultaneously exhibit low optical loss and sufficiently high refractive indices in the UV and EUV band. In addition, significant nanofabrication challenges further hinder progress, as the required sub-30-nm feature sizes and high–aspect-ratio nanostructures are approaching the practical limits of current lithography and etching technologies.

In this work, we propose a high-efficiency, polarization-insensitive, broadband achromatic metalens operating across the entire ultraviolet-C (UVC) band (200–280 nm). The proposed metalens is fabricated using $SiO_2$ and sapphire, materials that are friendly with space environments, thereby highlighting the potential of this platform for space-based application. The design methodology and structural details are described in Sec. 2, and the corresponding simulation results are discussed in Sec. 3.

## 2. Theory and method

To design a metalens, the required phase profile of the metalens must first be determined. As illustrated in Fig. 1(a), when a plane wave passes through the metalens, the transmitted light is focused at the designated focal point. This focusing behavior can be achieved if the phase imparted by the metalens satisfies the following equation[29]:

$$\phi(r, \omega) = -\frac{\omega}{c}\left(\sqrt{r^2 + F^2} - F\right) \tag{1}$$

where the c, $\omega$, r and F are the light speed, angular frequency, radial coordinate and focal length, respectively. In this design, the metalens has a diameter of D = 20 μm and operates across the 200–280 nm wavelength range, with a designed focal length of F = 200 μm. The phase profiles corresponding to different wavelengths are shown in Fig. 1(b). Using the phase profile at the central wavelength $\lambda_0$ = 240 nm as the reference $\phi_0$, the relative phase shifts at other wavelengths with respect to $\phi_0$ are illustrated in Fig. 1(c). Furthermore, by choosing the central wavelength $\lambda_0$ as the reference point, the required phase can be expressed as a Taylor series expansion, as given in Eq. (2):

$$\phi(r, \omega) = \phi(r, \omega_0) + \frac{\partial \phi}{\partial \omega}\bigg|_{\omega=\omega_0} (\omega - \omega_0) + \frac{\partial^2 \phi}{2\partial \omega^2}\bigg|_{\omega=\omega_0} (\omega - \omega_0)^2 + \ldots \tag{2}$$

where the central angular frequency $\omega_0 = 2\pi c / \lambda_0$. Eq. (2) implies that eliminating chromatic aberrations over a given bandwidth requires not only satisfying the zeroth-order term $\phi$, but also properly engineering the higher-order terms, which determine the dispersion characteristics. The first-order term $\partial\phi/\partial\omega$ and the second-order term $\partial2\phi/\partial\omega2$ are commonly referred to as the group delay (GD) and group delay dispersion (GDD), respectively. Their physical quantities typically lie in the femtosecond (fs) and femtosecond-squared (fs²) regimes in UVC band.

The term $\phi(r,\omega_0)$ represents the spherical wavefront required at the central frequency and determines whether light can be properly focused at the target focal plane for $\lambda0$, not providing any chromatic-dispersion compensation. GD describes the propagation time experienced by different frequency components as they pass through the metalens. A carefully engineered GD profile is essential to ensure that all wavelengths within the target spectral band arrive at the focal point simultaneously, thereby eliminating first-order chromatic aberration. GDD accounts for higher-order dispersion and characterizes how the GD varies with frequency, effectively describing the curvature of the dispersion profile. Proper control of GDD is critical for maintaining high-quality broadband imaging performance. Fig. 2 depicts the required GD profile as the function of radial coordinate for metalens with NA=0.05.

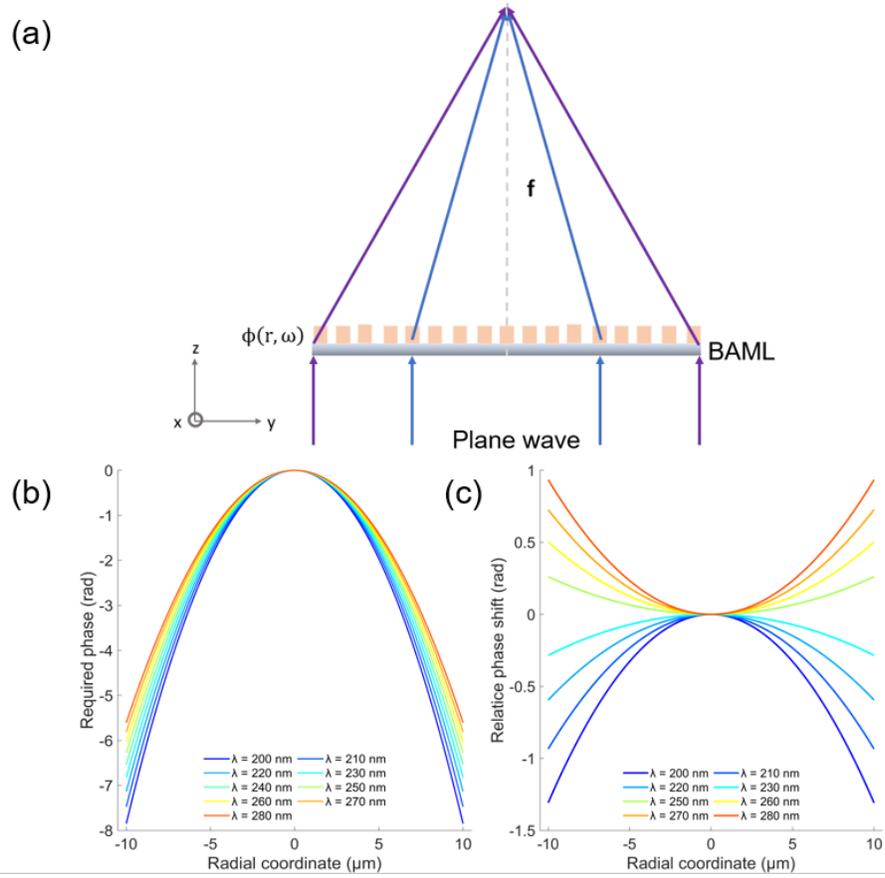

Fig. 1 (a) The schematic of BAML focuses the incident plane wave to a focal point with the designed focal length F. (b) The required phase distribution at different wavelengths as the function of radial coordinate for metalens and (c) relative phase shifts at other wavelengths with respect to the reference phase.

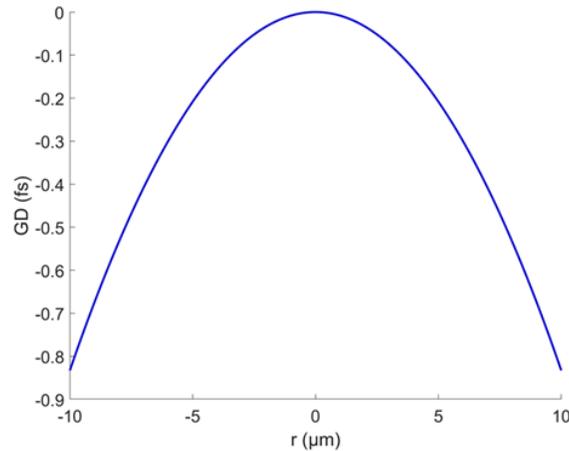

Fig. 2 The required group delay profile for the metalens coordinate.

As discussed in Sec. 2.1, the achromatic metalens must achieve the required phase profiles at different wavelengths in order to eliminate chromatic dispersion. We design a cross-shaped dielectric meta-atom composed of silicon dioxide ($SiO_2$), supported on a sapphire substrate and investigate whether this configuration can satisfy the above requirement. [2] The refractive index and extinction coefficient of sapphire and SiO2 are presented in Fig. 3. The schematic of the unit cell is shown in Fig. 4(a), where the side view depicts the height H and period P of the meta-atom, which are fixed at 400nm and 200nm respectively , and the incident light direction is indicated by the arrow. The corresponding top view is shown in Fig. 4(b), where the in-plane geometric parameters ($L_1$, $L_2$, $W_1$, $W_2$) are defined. Next, full-wave simulations were performed using the commercial software COMSOL Multiphysics to obtain the transmission amplitude and phase response of each meta-atom across the target wavelength range. Based on these results, we selected only the geometries exhibiting transmission efficiencies above 50% and computed their corresponding group delays (GD) then plug into Eq. 2 to construct the meta-atom library. The meta-atom whose phase responses closely match the required phase profile (Error < 5%) will be spatially arranged at their designated radial positions on the substrate to form the metalens. Once the layout is completed, the optical field distribution after light propagates through the metalens can be calculated.

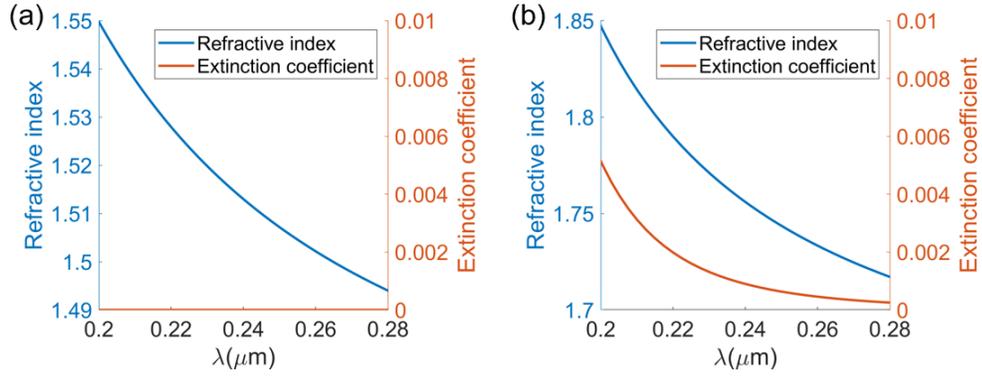

Fig. 3 The refractive index and extinction coefficient of (a) SiO2 and (b) sapphire.

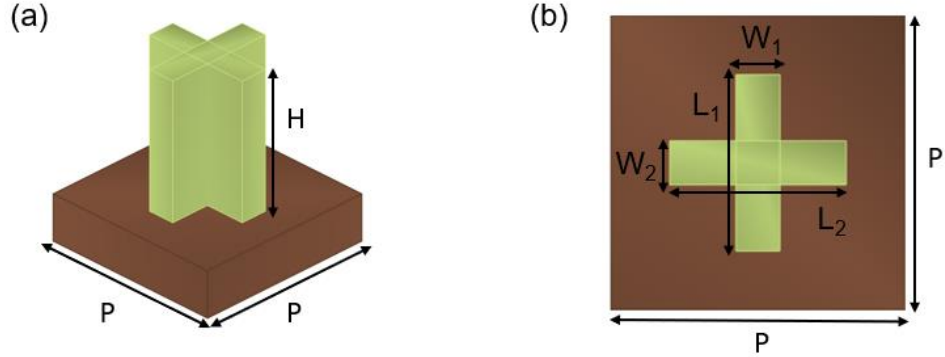

Fig. 4 (a) The side view of the cross-shaped meta-atom with H=400nm. The blue arrow represents the direction of the incident light. (b) The top view of the meta-atom.

## 3. Result and discussion

We selected three representative structures from the meta-atom library, as shown in Fig. 5(a), to illustrate their phase responses under illumination at different wavelengths. The results reveal a nearly linear phase variation with respect to frequency, enabling the extraction of the corresponding group delay . The transmittance of these structures are presented in Fig. 5(b) . Based on both the phase response and transmission efficiency, we can determine the realized phase profile for different wavelengths from meta-atom library compared to the required phase profile depicted in Fig. 6. Finally, the meta-atom geometries should be placed at specific radial positions to construct the metalens, as illustrated in Fig. 7(a) and Fig. 7(b).

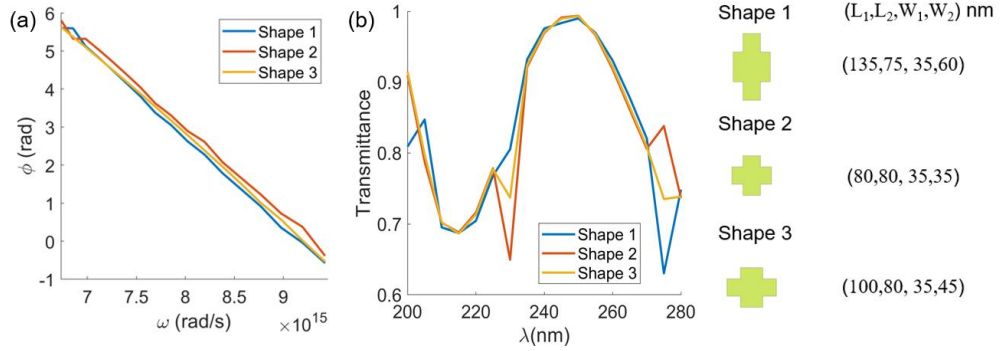

Fig. 5 Three representative meta-atom geometries are selected from the meta-atom library to demonstrate (a) their phase responses and (b) their corresponding transmittance spectra.

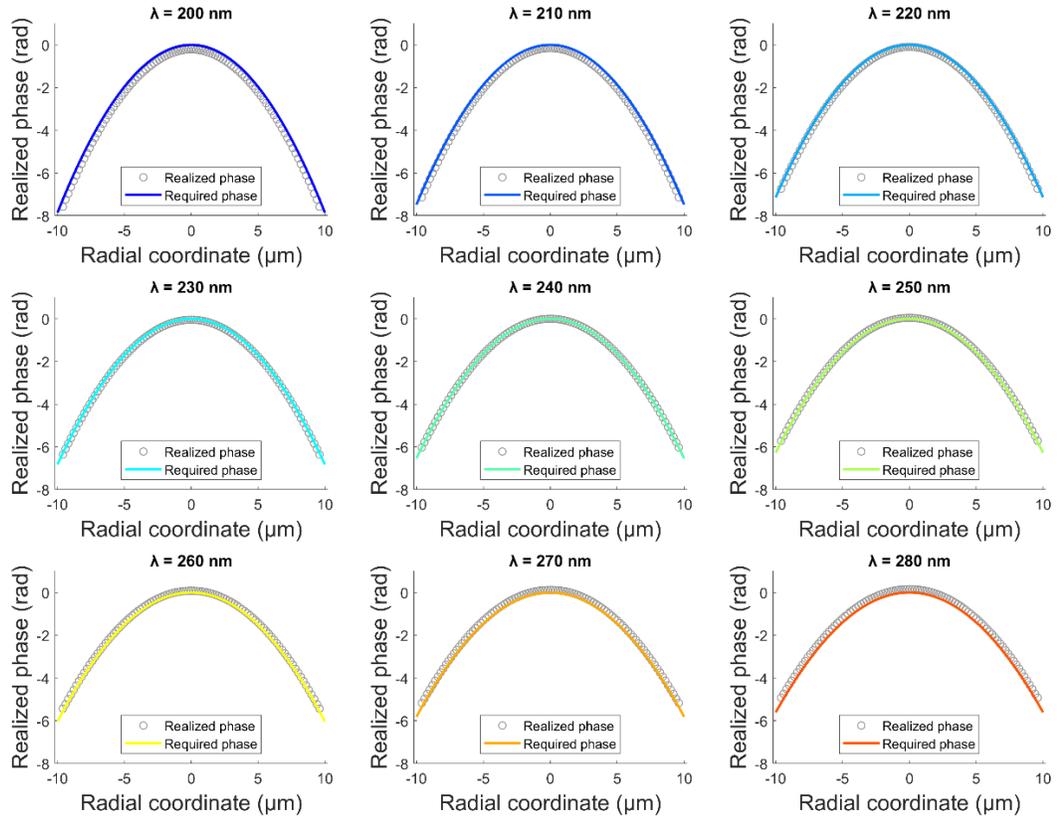

Fig. 6 The realized phase profile as function of metalens coordinate compared to the required phase.

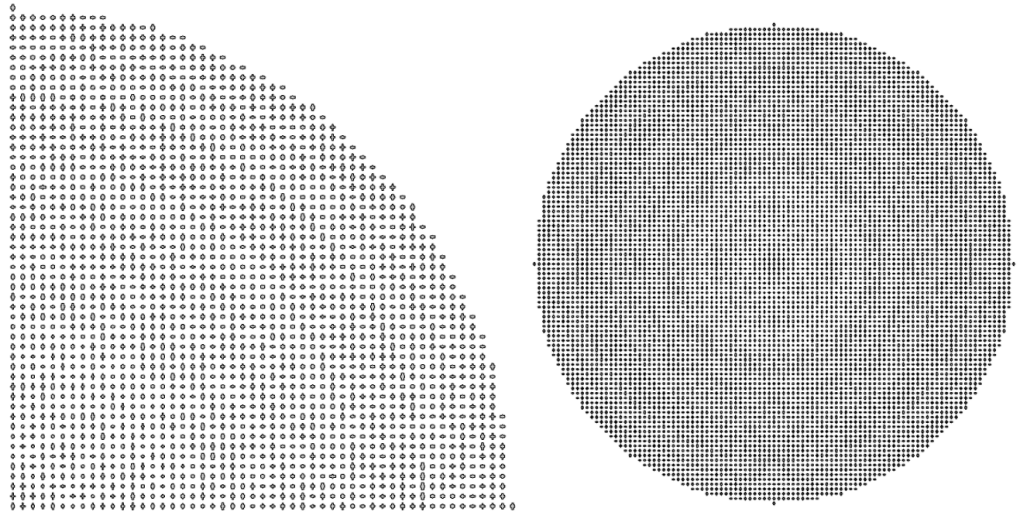

Fig. 7 Layout of the (a) quarter and the (b) full metalens.

Because the designed metalens exhibits cylindrical symmetry and is illuminated under normal incidence, the resulting far-field distribution is independent of the polarization state of the incident light . Therefore, we excite the metalens with a normally incident plane wave of arbitrary polarization and evaluate the transmitted optical field. For nine plane-wave wavelengths from 200 to 280 nm, the longitudinal intensity profile after propagation through the metalens is shown in Fig. 8, while the corresponding transverse normalized intensity distributions at z = 200μm plane and the intensity line profiles are presented in Fig. 9 . Fig. 5 –6 collectively demonstrate that the proposed metalens exhibits similar propagation characteristics across the entire UVC band and produces comparable beam sizes at the designed focal plane.

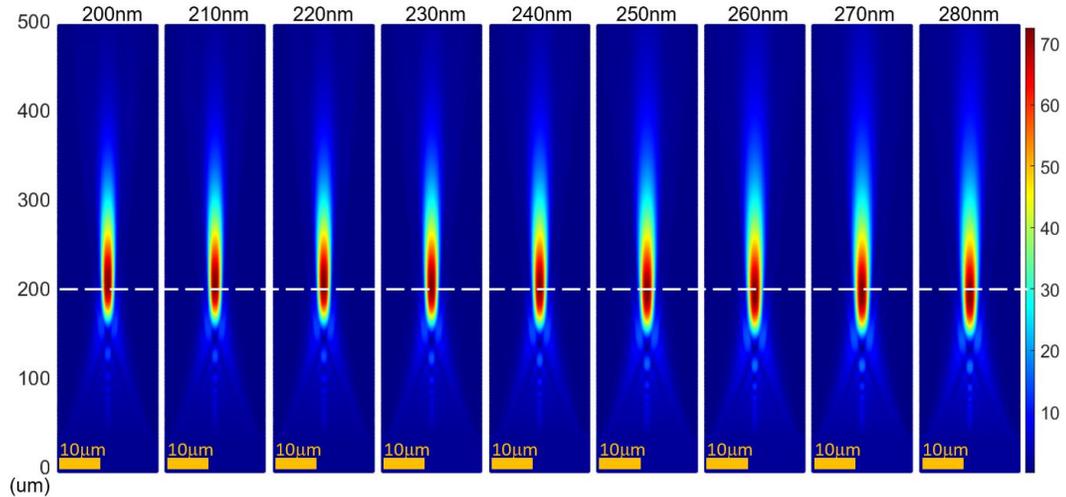

Fig. 8 The longitudinal intensity profile after the incident light passes through the metalens. Scale bar, 10um. The white dash line indicates the focal plane at z=200um corresponding to the designed focal plane.

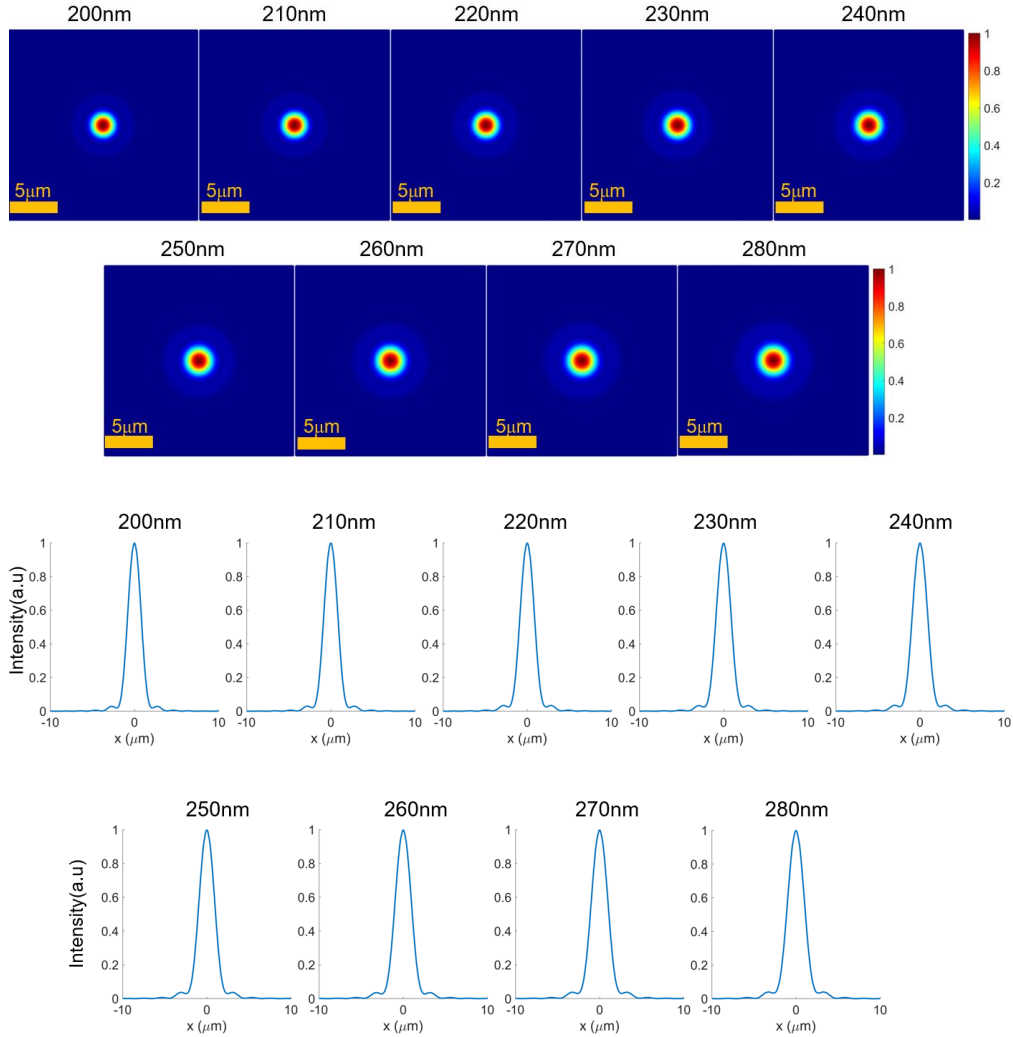

Fig. 9 The transverse intensity profile and intensity line profile at designed focal plane under different wavelengths.

The focal positions and focusing efficiencies at different wavelengths are summarized in Fig. 10(a), while the corresponding beam radius are shown in Fig. 10(b). the focal position is defined as the axial location where the on-axis intensity reaches its maximum, and the focusing efficiency is defined as the ratio between the optical power in the focal plane and the incident power on the metalens aperture. The beam radius is defined as the radial position where the intensity drops to 1/e2 of the on-axis peak value. The results indicate that the focal shift relative to the designed focal length is small , confirming good achromatic performance over the target bandwidth. Here, the focal shift is defined as 100% x (focal position - F)/F. The largest focal shift is about 10% at 210nm by Fig. 10(a). In addition, the focusing efficiency exceeds 75% across all wavelengths, implying that there is no significant energy loss into unwanted diffraction channels.  These characteristics demonstrate that the proposed design successfully realizes a broadband achromatic metalens, which is promising for applications in spaceborne observation and compact imaging systems.

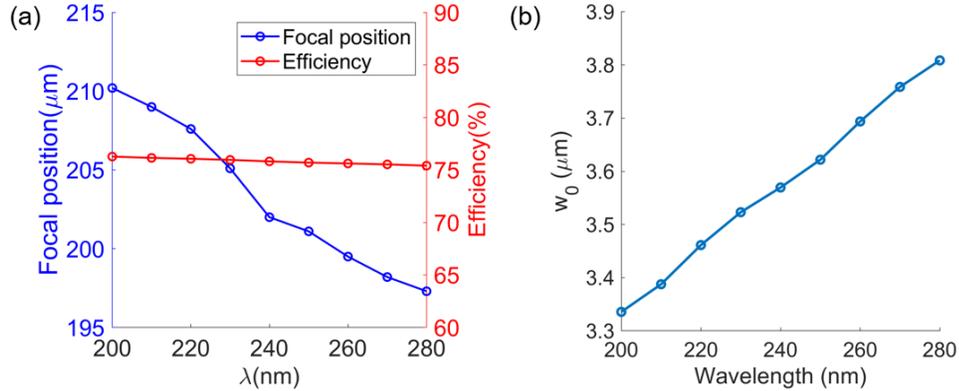

Fig. 10 (a) The focal position, efficiency and (b) beam radius under different wavelengths.

## 4. Conclusion

In summary, this paper successfully design dispersion-engineered achromatic metalens (NA~0.05) across the UVC band with a small focal shift , polarization independence and high focusing efficiency more than 75%. By properly adjusting the geometric parameters of the SiO2 meta-atom, the phase and group delay library will be build then we can construct the full metalens. This research holds great potential for applications in spaceborne observation and compact imaging systems.

**Funding.** Ministry of Science and Technology, Taiwan (NSTC 112-2112-M-006-018 -).

**Acknowledgments.**

**Disclosures.** The authors declare no conflicts of interest.

**Data Availability.** Data will be made available on request.

**References**

1. N. Yu, P. Genevet, M.A. Kats, F. Aieta, J.-P. Tetienne, F. Capasso, Z. Gaburro, Light propagation with phase discontinuities: generalized laws of reflection and refraction, Science 334 (2011) 333–337.
2. W.T. Chen, A.Y. Zhu, F. Capasso, Flat optics with dispersion-engineered metasurfaces, Nat. Rev. Mater. 5 (2020) 604–620.
3. X.G. Luo, Principles of electromagnetic waves in metasurfaces, Sci. China Phys. Mech. Astron. 58 (2015) 594201.
4. A.H. Dorrah, F. Capasso, Tunable structured light with flat optics, Science 376 (2022) 6591.
5. H.-H. Hsiao, C.H. Chu, D.P. Tsai, Fundamentals and applications of metasurfaces, Small Methods 1 (2017) 1600064.
6. M. Pan, Y. Fu, M. Zheng, et al. Dielectric metalens for miniaturized imaging systems: progress and challenges. Light Sci Appl 11, 195 (2022).
7. T. Li, C. Chen, X. Xiao, J. Chen, S. Hu, S. Zhu, Revolutionary meta-imaging: from superlens to metalens, Photon. Insights 2 (2023) R01.
8. X. Zou, G. Zheng, Q. Yuan, W. Zang, R. Chen, T. Li, L. Li, S. Wang, Z. Wang, S. Zhu, Imaging based on metalenses, Photonix 1 (2020) 2.
9. M. Khorasaninejad, W.T. Chen, R.C. Devlin, J. Oh, A.Y. Zhu, F. Capasso, Metalenses at visible wavelengths: diffraction-limited focusing and subwavelength resolution imaging, Science 352 (2016) 1190–1194.
10. M. Khorasaninejad, W.T. Chen, A.Y. Zhu, J. Oh, R.C. Devlin, D. Rousso, F. Capasso, Multispectral chiral imaging with a metalens, Nano Lett. 16 (2016) 4595–4600.
11. Z. Huang, Y. Zheng, J. Li, Y. Cheng, J. Wang, Z.-K. Zhou, L. Chen, High-resolution metalens imaging polarimetry, Nano Lett. 23 (2023) 10991–10997.

12. W.T. Chen, K.-Y. Yang, C.-M. Wang, Y.-W. Huang, G. Sun, I.-D. Chiang, C.Y. Liao, W.-L. Hsu, H.-T. Lin, S. Sun, L. Zhou, A.Q. Liu, D.P. Tsai, High-efficiency broadband meta-hologram with polarization-controlled dual images, Nano Lett. 14 (2014) 225–230.
13. S.C. Malek, H.-S. Ee, R. Agarwal, Strain multiplexed metasurface holograms on a stretchable substrate, Nano Lett. 17 (2017) 3641–3645.
14. J. He, T. Dong, B. Chi, S. Wang, X. Wang, Y. Zhang, Meta-hologram for three-dimensional display in terahertz waveband, Microelectron. Eng. 220 (2020) 111151.
15. S. Wan, C. Dai, Z. Li, L. Deng, Y. Shi, W. Hu, G. Zheng, S. Zhang, Z. Li, Toward water-immersion programmable meta-display, Adv. Sci. 9 (2022) 22025581.
16. Z. Li, Y. Shi, C. Dai, Z. Li, On-chip-driven multicolor 3D meta-display, Laser Photonics Rev. 18 (2024) 2301240.
17. T. Stone, N. George, Hybrid diffractive–refractive lenses and achromats, Appl. Opt. 27 (1988) 2960–2971.
18. Pin Chieh Wu, Vin-Cent Su, Yi-Chieh Lai, Cheng Hung Chu, Jia-Wern Chen, Shen-Hung Lu, Ji Chen, Beibei Xu, Chieh-Hsiung Kuan, Tao Li, Shining Zhu, Din Ping Tsai, Broadband achromatic optical metasurface devices, Nat. Commun. 8, 187 (2017).
19. Y. Xie, J. Zhang, S. Wang, D. Liu, and X. Wu, Broadband polarization-insensitive metalens integrated with a charge-coupled device in the short-wave near-infrared range, Opt. Express 30 (7) (2022) 11372.
20. Y. Zheng, W. Zhu, L. Xia, G. Chen, Y. Li, S. Dang, M. Zhang, C. Du, Achromatic metalens in visible band based on nano double trapezoid structures, Opt. Mater. 161 (2025) 116801.
21. Z. Huang, M. Qin, X. Guo, C. Yang, S. Li, Achromatic and wide-field metalens in the visible region, Opt. Express 29 (2021) 13542–13551.
22. Y. Wang, Q. Chen, W. Yang, Z. Ji, L. Jin, X. Ma, Q. Song, A. Boltasseva, J. Han, V. M. Shalaev, S. Xiao, High-efficiency broadband achromatic metalens for near-IR biological imaging window, Nat. Commun. 12, 5560 (2021).
23. S. Yue, Y. Liu, R. Wang, Y. Hou, H. Shi, Y. Feng, Z. Wen, Z. Zhang, All-silicon polarization-independent broadband achromatic metalens designed for the mid-wave and long-wave infrared, Opt. Express 31, 44340–44352 (2023).
24. Fan Xu, Wenjie Chen, Ming Li, Peng Liu, Yuhang Chen, Broadband achromatic and wide field-of-view single-layer metalenses in the mid-infrared, Opt. Express 31(22), 36439–36450 (2023).
25. Z. Zeng, X. Chen, L. Du, J. Li, and L. Zhu, Design of a dielectric ultrathin near-infrared metalens based on electromagnetically induced transparency, Opt. Mater. Express 13 (9) (2023) 2541–2549.
26. J. Kim, Y. Kim, W. Kim, D. Oh, D. Kang, J. Seong, J. W. Shin, D. Go, C. Park, H. Song, J. Ahn, H. Lee, J. Rho, 8″ wafer-scale, centimeter-sized, high-efficiency metalenses in the ultraviolet, Mater. Today 73 (2024) 9–15.
27. K. Cheng, H. Cui, Q. Li, Y. Zhao, Y. Zhou, Simulation of multiwavelength achromatic metalens in the extreme ultraviolet, Opt. Commun. 557 (2024) 130345.
28. F. Ali and S. Aksu, A hybrid broadband metalens operating at ultraviolet frequencies, Sci. Rep. 11 (2021) 2303.
29. W. T. Chen, A. Y. Zhu, V. Sanjeev, M. Khorasaninejad, Z. Shi, E. Lee, and F. Capasso, A broadband achromatic metalens for focusing and imaging in the visible, Nat. Nanotechnol. 13 (2018) 220–226.